%% file: main.tex
\UseRawInputEncoding
\pdfoutput=1 
\documentclass{nature}

\usepackage{multirow}
\usepackage{ccaption}
\usepackage{subcaption}
\usepackage{graphicx}
\usepackage{lipsum}
\usepackage{ragged2e}
\usepackage{fixltx2e}
\usepackage{hyperref}

\RequirePackage{mathpazo}
\usepackage{amsmath}

\usepackage{comment}
\usepackage{cite}
\usepackage{adjustbox} 

\makeatletter
\newenvironment{figurehere}
{\def\@captype{figure}}
{}
\makeatletter

\usepackage{xcolor}

\title{\begin{center}  Powering Disturb-Free Reconfigurable Computing \\ 
and Tunable Analog Electronics with Dual-Port Ferroelectric FET\end{center}}

\author
{ \centering
{
\large{Zijian Zhao$^{1}$, Shan Deng$^{1}$, Swetaki Chatterjee$^{2,6}$, Zhouhang Jiang$^{1}$, Muhammad Shaffatul Islam$^{1}$, Yi Xiao$^{3}$, Yixin Xu$^{3}$, Scott Meninger$^{4}$, Mohamed Mohamed$^{4}$, Rajiv Joshi$^{5}$, Yogesh Singh Chauhan$^{6}$, \\
Halid Mulaosmanovic$^{7}$, Stefan Duenkel$^{7}$, Dominik Kleimaier$^{7}$, Sven Beyer$^{7}$, Hussam Amrouch$^{8,9}$, Vijaykrishnan Narayanan$^{3}$, and Kai Ni$^{1\dagger}$\\}
\vspace{3ex}
\normalsize{$^{1}$Rochester Institute of Technology, Rochester, NY 14623, USA}\\
\normalsize{$^{2}$University of Stuttgart, Stuttgart, Germany}\\
\normalsize{$^{3}$Pennsylvania State University, State College, PA 16802, USA}\\
\normalsize{$^{4}$MIT Lincoln Laboratory, Lexington, MA 02421,USA}\\
\normalsize{$^{5}$IBM Thomas J. Watson Research Center, USA}\\
\normalsize{$^{6}$IIT Kanpur, Kanpur, India}\\
\normalsize{$^{7}$GlobalFoundries Fab1 LLC \& Co. KG, Dresden, Germany}\\
\normalsize{$^{8}$Technical University of Munich (TUM), Munich, Germany}\\
\normalsize{$^{9}$Munich Institute of Robotics and Machine Intelligence, Munich, Germany}\\
\vspace{2ex}
\normalsize{$^{\dagger}$To whom correspondence should be addressed} \\
\normalsize{Email: kai.ni@rit.edu} \\
}}

\begin{document}
\flushbottom
\maketitle
\vspace{3ex}
\begin{abstract}
Single-port ferroelectric FET (FeFET) that performs write and read operations on the same electrical gate prevents its wide application in tunable analog electronics and suffers from read disturb, especially to the high-threshold voltage (\textit{V}\textsubscript{TH}) state as the retention energy barrier is reduced by the applied read bias. To address both issues, we propose to adopt a read disturb-free dual-port FeFET where write is performed on the gate featuring a ferroelectric layer and the read is done on a separate gate featuring a non-ferroelectric dielectric. Combining the unique structure and the separate read gate, read disturb is eliminated as the applied field is aligned with polarization in the high-\textit{V}\textsubscript{TH} state and thus improving its stability, while it is screened by the channel inversion charge and exerts no negative impact on the low-\textit{V}\textsubscript{TH} state stability. Comprehensive theoretical and experimental validation have been performed on fully-depleted silicon-on-insulator (FDSOI) FeFETs integrated on 22 nm platform, which intrinsically has dual ports with its buried oxide layer acting as the non-ferroelectric dielectric. Novel applications that can exploit the proposed dual-port FeFET are proposed and experimentally demonstrated for the first time, including FPGA that harnesses its read disturb-free feature and tunable analog electronics (e.g., frequency tunable ring oscillator in this work) leveraging the separated write and read paths.
\end{abstract}

%
%
\thispagestyle{empty}


\input{bibtex/sections/1_Introduction}

\input{bibtex/sections/2_Operation_principles}

\input{bibtex/sections/3_Experimental_verification}

\input{bibtex/sections/4_Applications}
\input{bibtex/sections/5_Conclusion}

\section*{\large Data availability}
The data that support the plots within this paper and other findings of this study are available from the corresponding author on reasonable request.

\section*{\large References}
\bibliography{ref1.bib}

\section*{\large Acknowledgements}

This work was partly supported by SUPREME and PRISM, two of the JUMP 2.0 centers, Army Research Office under Grant Number W911NF-21-1-0341, NSF 2008365, and the United States Air Force under Air Force Contract No. FA8702-15-D-0001. Any opinions, findings, conclusions or recommendations expressed in this material are those of the author(s) and do not necessarily reflect the views of the United States Air Force. The device fabrication was supported by the German Bundesministerium für Wirtschaft (BMWI) and by the State of Saxony in the frame of the “Important Project of Common European Interest (IPCEI: WIN-FDSOI).”

\section*{\large Author contributions}

K.N. and V.N., proposed and supervised the project. S.C., M.I., Y.C., and H.A. conducted device simulations. Y.X. and Y.X. conducted FPGA design. Z.Z., S.D., and Z.J. performed experimental verification. M.M. and S.M. conducted the ring oscillator design and simulation. H.M., S.D., D.K., and S.B. fabricated the FeFET devices. R.J. helped with the device validation. All authors contributed to write up of the manuscript. 

\section*{\large Competing interests}
The authors declare no competing interests.

\newpage

\input{bibtex/sections/Supplementary}

\end{document}

%% file: bibtex/sections/1_Introduction.tex
\section*{\large Introduction}

HfO\textsubscript{2} based FeFET is a prime candidate for embedded nonvolatile memory (eNVM) due to its excellent CMOS compatibility, scalability, and energy efficiency \cite{salahuddin2018era,khan2020future,mikolajick2020past,mulaosmanovic2021ferroelectric,schroeder2022fundamentals}. Ever since the discovery of ferroelectricity in doped HfO\textsubscript{2}, significant progress has been made in its integration in advanced technology platforms, thus realizing scaled planar \cite{trentzsch201628nm,dunkel2017fefet}, FinFET \cite{bae2020sub,yan2021high}, and gate-all-around FeFETs \cite{liao2022endurance,huang2021ferroelectric}. The more the number of physical gates, the better the electrostatic gate control over the channel. However, irrespective of the number of physical gates present, it typically has only a single electrical port, which is utilized for both read and write operations, as shown in Fig. \ref{fig:Fig1}\textbf{c-d}. This introduces potential read disturb to the high-\textit{V}\textsubscript{TH} state. As shown in Fig.\ref{fig:Fig1}\textbf{c}, a negative write pulse applied on the gate switches the polarization to point towards the gate metal, which sets the FeFET to the high-\textit{V}\textsubscript{TH} state. During retention after the write pulse, ideally the two polarization states are separated by an energy barrier, \textit{E}\textsubscript{B}, which determines the state stability, as shown in Fig.\ref{fig:Fig1}\textbf{e}. During the read operation, a read pulse of magnitude, \textit{V}\textsubscript{READ}, which is between the low-\textit{V}\textsubscript{TH} state and the high-\textit{V}\textsubscript{TH} state is applied to sense the polarization states. This read bias, however, lowers the energy barrier that prevents the high-\textit{V}\textsubscript{TH} state from flipping to the low-\textit{V}\textsubscript{TH} state, as shown in Fig.\ref{fig:Fig1}\textbf{e}, to $E_{B}^{'}$, thus bringing read disturb issue.         

This can be further understood from the switching dynamics in a HfO\textsubscript{2} thin film, as shown in Fig.\ref{fig:Fig1}\textbf{f}, which shows a typical dependence between the switching time and the applied voltage for the high-\textit{V}\textsubscript{TH} state to switch to the low-\textit{V}\textsubscript{TH} state. In a thin poly-crystalline HfO\textsubscript{2} thin film, the polarization switching is believed to follow the nucleation-limited switching model, where the rate-limiting process is for the multiple domains to nucleate locally\cite{mulaosmanovic2017switching}. The required switching time is an exponential function of applied pulse amplitude such that during the write operation, the high electric field renders an ultra-fast polarization switching (i.e., $\leq$ns) while during the retention, the small electric field can yield a necessarily long memory retention time (i.e., $\geq$10 years). Following the same dynamics, at \textit{V}\textsubscript{READ} (i.e., typically around 1 V), the required switching time can be much less than the retention time. This read disturb issue might not be a big concern for FeFET to operate as a memory because repetitive reading of the same FeFET is highly unlikely. However, there are applications that need to apply the \textit{V}\textsubscript{READ} constantly, which brings serious concern over the state stability. A noteworthy example is FeFET based reconfigurable computing, where the FeFET is exploited as a compact nonvolatile switch as shown in Fig.\ref{fig:Fig1}\textbf{a} \cite{yu2022hardware,xu2022ferroelectric}. By combining the configuration memory and a switch together, a FeFET based nonvolatile switch can implement the routing network in a field programmable gate array (FPGA) by constructing the connection box (CB) and switch box (SB). In this application, to activate the written configuration, the \textit{V}\textsubscript{READ} has to be applied constantly. The read disturb issue poses a serious challenge for this application. 

\begin{figurehere}
  \centering
    \includegraphics[scale=0.9,width=\textwidth]{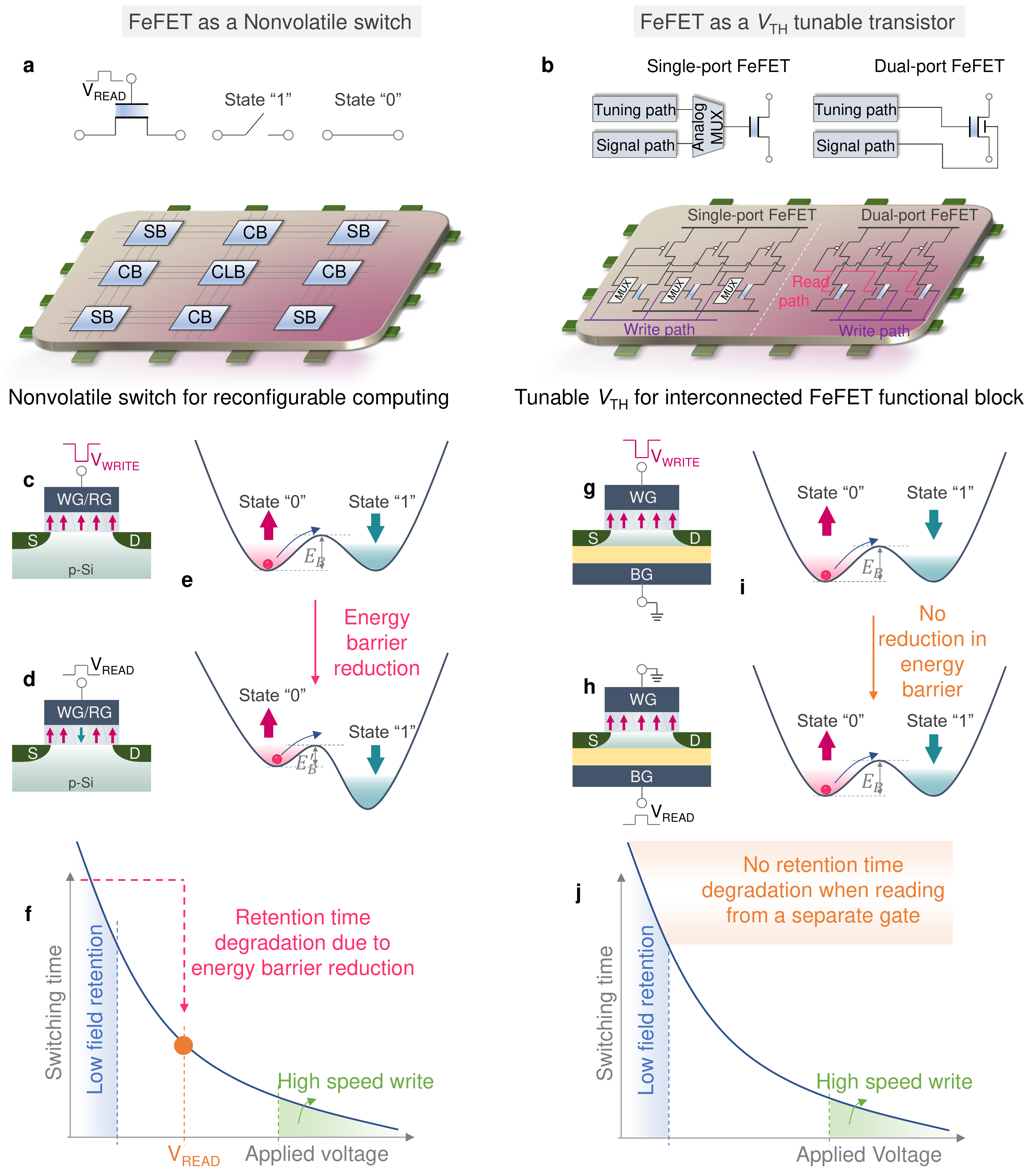}
    \captionsetup{parbox=none} 
    \caption{\textbf{Dual-port FeFET for read disturb-free operation and separated write and read path operation.} \textbf{a.} A single-port FeFET can function as a nonvolatile switch for reconfigurable computing, e.g., FPGA. The SB, CB, and LUT in the CLB leverage FeFETs to implement reconfigurable logic with a high density. \textbf{b.} Functional blocks consist of \textit{V}\textsubscript{TH} tunable single-port FeFETs and analog MUXs for write/read switching, which increases design complexity. Dual-port FeFET offers separated write and read paths, which simplify the design of functional blocks by eliminating analog MUXs. \textbf{c-d.} In the case of the FeFET written to high-\textit{V}\textsubscript{TH} state, constant read operation will cause polarization switching. \textbf{e.} Energy landscape illustrating the energy barrier to prevent polarization switching in the high-\textit{V}\textsubscript{TH} state is reduced by the applied read gate bias. \textbf{f.} Retention time is degraded due to reduced energy barrier as \textit{V}\textsubscript{READ} increases. \textbf{g-h.} The separated read and write gates can retain ferroelectric polarization for the high-\textit{V}\textsubscript{TH} state. \textbf{i.} Energy landscape demonstrating energy barrier to prevent polarization flip in a high-\textit{V}\textsubscript{TH} state remains unchanged when sensing from a read gate in a dual-port FeFET. \textbf{j.} Retention degradation caused by the read operation is eliminated. }
    \label{fig:Fig1}
\end{figurehere}

In addition to the read disturb issue, the voltage supply poses an overhead for a single-port FeFET sharing the read and write port. Since the write voltage and the read voltage are usually different, there is a need of analog multiplexer that can switch between the two. For applications like tunable analog electronics that try to exploit FeFET for its programmable \textit{V}\textsubscript{TH} and intrinsic transistor structure, the need to switch between the signal path and the tuning path necessitates an analog multiplexer for each FeFET,  diminishing the benefits of adopting FeFETs for those applications. Fig.\ref{fig:Fig1}\textbf{b} shows an example of ring oscillator with FeFETs in the pull down path, which can tune the oscillation frequency. Since each FeFET receives a different signal, that prevents sharing of analog multiplexer among single-port FeFETs, like in a memory array. To address both issues, this work proposes to adopt dual-port FeFETs with separate write and read paths and demonstrates its applications in the aforementioned two applications.  

In a dual-port FeFET, two electrical gates are available where one gate that features a ferroelectric thin film is dedicated for the write operation (Fig.\ref{fig:Fig1}\textbf{g}) and the other gate with a non-ferroelectric dielectric thin film is used for the read operation (Fig.\ref{fig:Fig1}\textbf{h}). For a dual-port FeFET with an ultra-thin channel, such as the FDSOI FeFET, the polarization programmed through the write gate modulates the channel carrier concentration, which can be also read out through the read gate leveraging the coupling of the two gates. The dual-port concept has been proposed a decade ago\cite{kaneko2011dual,jang2013self,ahn2016dual} and has been revived recently on FeFET \cite{jiang2022asymmetric, mulaosmanovic2021ferroelectric2}. This structure shows two interesting properties. First, by properly designing a thick non-ferroelectric dielectric layer, sensing through the read gate can amplify the memory window compared with that sensed through the write gate \cite{mulaosmanovic2021ferroelectric2,jiang2022asymmetric}. The memory window amplification is interesting, for example 4 V write voltage can induce 10 V memory window\cite{mulaosmanovic2021ferroelectric2,jiang2022asymmetric}, and it needs further studies on the possibility of enabling low-voltage high density memory~\cite{chatterjee2022ted}. Second, sensing through the read gate enables read disturb-free feature, which has been suggested in multiple reports but has never been substantiated\cite{kaneko2011dual,mulaosmanovic2021ferroelectric2}. Therefore, in this work, we will provide combined theoretical and experimental validation of this property and firmly establish its physical ground.

With our proposed structure, the two problems related to single-port FeFET are easily addressed. As shown in Fig.\ref{fig:Fig1}\textbf{i}, after a memory write, an energy barrier, \textit{E}\textsubscript{B}, separating the two polarization states remains unchanged or even gets strengthened when sensing through the read gate. In this way, the retention time is not degraded, as shown in Fig.\ref{fig:Fig1}\textbf{j}. Except for the reconfigurable computing application mentioned here, the disturb-free feature could also find applications in the high density vertical NAND-type storage to mitigate the write/read disturb introduced on unselected pages when a high pass voltage needs to be applied for them to pass biases\cite{compagnoni2019reliability}. This can greatly improve the vertical NAND array reliability. In addition, the separation between the write and read paths opens up the application space of FeFETs for tunable analog electronics by embedding dual-port FeFETs in circuits (see Fig.\ref{fig:Fig1}\textbf{b}), where the signal is sent to the read gate and the tuning pulse is forwarded to the write gate. Therefore, the analog multiplexer is avoided, greatly simplifying the supporting circuitry.

The contributions of this work are: i) establishing theoretical understanding on the origin of read disturb-free operation in the dual-port FeFETs; ii) performing extensive experimental validation of the read disturb-free operation on a dual-port FeFET; iii) exploiting the read disturb-free feature in the dual-port FeFET to build a FPGA by demonstrating the primitives, including the look up table (LUT) and routing elements; iv) leveraging the separated write and read paths in the dual-port FeFETs to open up the application space of FeFETs for tunable analog electronics (e.g., frequency tunable ring oscillator in this work), which has long been neglected for single-port FeFET due to the necessity of complex analog multiplex between signal path and tuning path.

%% file: bibtex/sections/2_Operation_principles.tex
\section*{\large Operation Principles of Read Disturb-Free Operation}

The origin of the read disturb-free operation in a dual-port FeFET is presented and compared with single-port FeFET in Fig.\ref{fig:Fig2}. When the FeFET is programmed to the low-\textit{V}\textsubscript{TH} state, applying a read pulse on the write gate counteracts the depolarization field, \textit{E}\textsubscript{Dep}, thus enhancing the state stability, as shown in Fig.\ref{fig:Fig2}\textbf{a}. For a FeFET programmed to the high-\textit{V}\textsubscript{TH} state, the \textit{E}\textsubscript{Dep} is pointing towards the channel. When a positive read voltage is applied, the exerted electric field, \textit{E}\textsubscript{App}, reinforces the \textit{E}\textsubscript{Dep}, thus creating concerns over the state stability as shown in Fig.\ref{fig:Fig2}\textbf{b}. If a long-term read bias, \textit{V}\textsubscript{READ}, has to be applied, such as the routing switch in a FPGA, this can result in performance degradation or even a function failure. This issue is addressed in a dual-port FeFET. As seen in Fig.\ref{fig:Fig2}\textbf{c}, when reading a FeFET in the low-\textit{V}\textsubscript{TH} state through the read gate, the applied electric field is screened by the channel and most of the read voltage is dropped across the non-FE dielectric. The electric field in the ferroelectric remains intact and therefore no degradation in retention is expected. Additionally, when the FeFET is configured to the high-\textit{V}\textsubscript{TH} state, there is no channel formed in the semiconductor and therefore no screening happens. However, the \textit{E}\textsubscript{App} counteracts the \textit{E}\textsubscript{Dep} as shown in Fig.\ref{fig:Fig2}\textbf{d}, therefore improvement, rather than degradation, in retention is expected. 

\begin{figurehere}
  \centering
    \includegraphics[scale=1,width=1\textwidth]{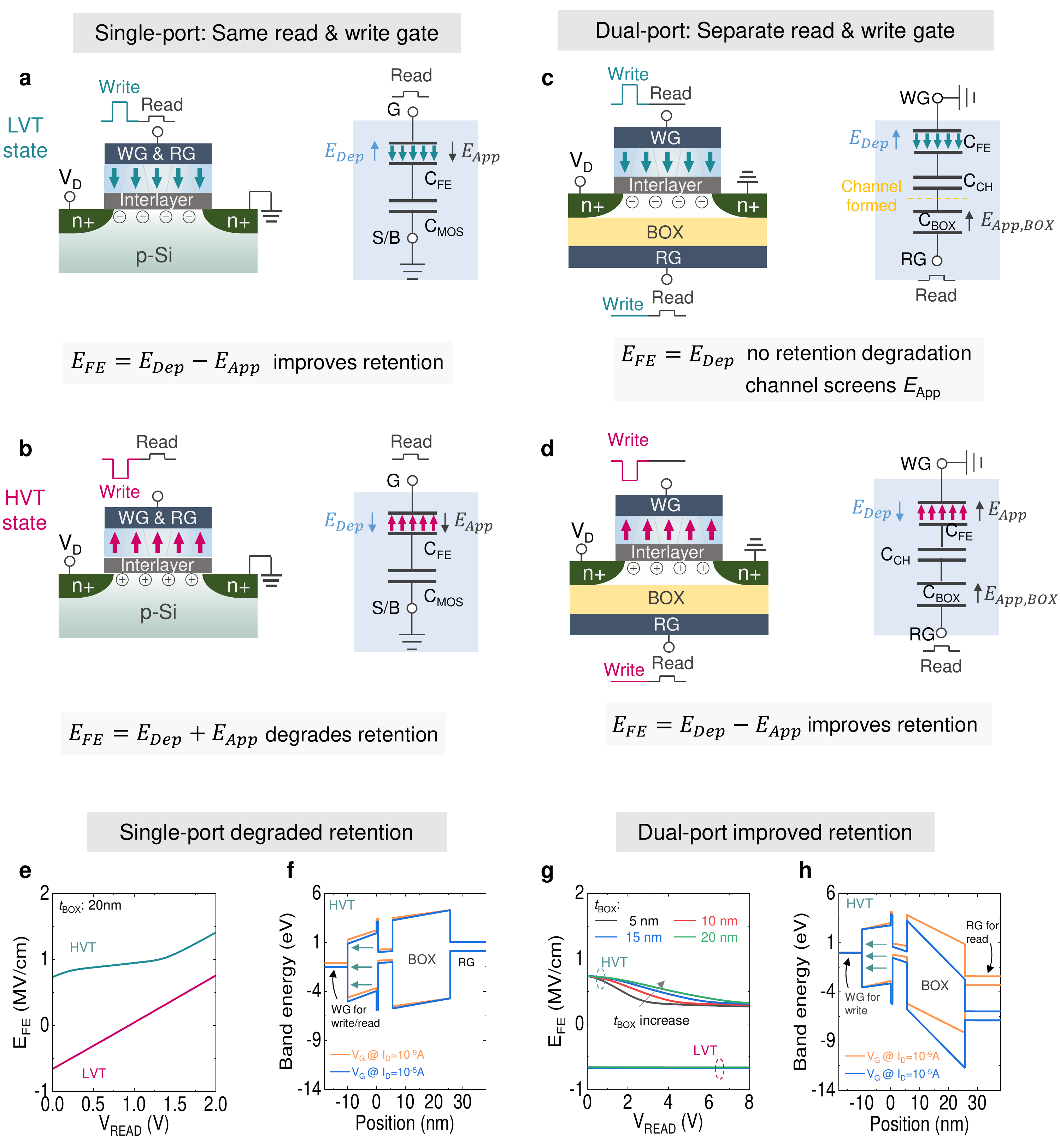}
    \captionsetup{parbox=none} 
    \caption{\textbf{Origin of read disturb-free operation in a dual-port FeFET.} \textbf{a.} For a single-port FeFET written to the low-\textit{V}\textsubscript{TH} state, the \textit{E}\textsubscript{App} exerted by the read bias reduces \textit{E}\textsubscript{Dep}, thus improving retention. \textbf{b.} For a single-port FeFET written to the high-\textit{V}\textsubscript{TH} state, the \textit{E}\textsubscript{App} exerted by the read bias reinforces \textit{E}\textsubscript{Dep}, thus degrading retention. \textbf{c.} In a dual-port FeFET, channel formation screens \textit{E}\textsubscript{App}, retaining low-\textit{V}\textsubscript{TH} state retention. \textbf{d.} \textit{E}\textsubscript{App} counteracts \textit{E}\textsubscript{Dep} in the high-\textit{V}\textsubscript{TH} state, thus improving retention. \textbf{e.} The relationship between \textit{E}\textsubscript{FE} and \textit{V}\textsubscript{READ} for the single-port FeFET confirms the analysis of the read disturb issue. \textbf{f.} TCAD simulation of band diagram across the stack at two different read biases. \textbf{g.} The relationship between \textit{E}\textsubscript{FE} and \textit{V}\textsubscript{READ} for the dual-port FeFET confirms the origin of the read disturb-free operation and the feasibility of scaling non-FE layer. \textbf{h.} Band diagram across the gate stack shows the majority of \textit{V}\textsubscript{READ} drops across non-FE dielectric.}
    \label{fig:Fig2}
\end{figurehere}

To validate the operation principles, simulations using Sentaurus technology computer aided design (TCAD) tool is conducted. A FDSOI FeFET TCAD model is built by adapting a calibrated FDSOI logic transistor with experimentally measured \textit{I}\textsubscript{D}-\textit{V}\textsubscript{G} characteristics\cite{ni2021channel}. First the single-port FeFET operation is studied. Fig.\ref{fig:Fig2}\textbf{e} shows the ferroelectric electric field, \textit{E}\textsubscript{FE}, as a function of the \textit{V}\textsubscript{READ} applied on the same write gate for both low-/high-\textit{V}\textsubscript{TH} states. It shows that the electric field that is against the polarization is reduced/enhanced with the increasing \textit{V}\textsubscript{READ} for low-/high-\textit{V}\textsubscript{TH} states, respectively, confirming the proposed working principles. Fig.\ref{fig:Fig2}\textbf{f} further shows the band diagrams across the stack for the \textit{I}\textsubscript{D} at 1 nA and 10 $\mu$A for high-\textit{V}\textsubscript{TH} state sensing. It also confirms more electric field drop in the ferroelectric at a larger \textit{V}\textsubscript{READ}. However, when sensing from a separate read gate, the read disturb can be avoided. Fig.\ref{fig:Fig2}\textbf{g} shows the \textit{E}\textsubscript{FE} at different \textit{V}\textsubscript{READ} at the read gate. It shows that the \textit{E}\textsubscript{FE} against the polarization is constant/reduced with the \textit{V}\textsubscript{READ} for the low-/high-\textit{V}\textsubscript{TH} states, which will lead an non-degraded/enhanced retention, respectively. The band diagrams for the high-\textit{V}\textsubscript{TH} state in a dual-port FeFET shown in Fig.\ref{fig:Fig2}\textbf{h} also show that most of the \textit{V}\textsubscript{READ} is dropped across the non-FE dielectric. The FE depolarization field is counteracted partially, therefore enhancing the state stability. For completeness, the band diagrams for sensing the low-\textit{V}\textsubscript{TH} state in both single-port and dual-port FeFETs  are presented in supplementary Fig.\ref{fig:sups1}. It agrees with the working principles presented. These results confirm that having a dedicated non-FE read gate can enable the read disturb-free operation. 

It is also worth noting that the read disturb-free operation is not dependent on a particular non-FE dielectric used. Fig.\ref{fig:Fig2}\textbf{g} shows the \textit{E}\textsubscript{FE} for different SiO\textsubscript{2} thicknesses, ranging from 5 nm to 20 nm, when sensing a dual-port FeFET. It suggests that there is no change in the \textit{E}\textsubscript{FE} for different SiO\textsubscript{2} thicknesses when sensing a low-\textit{V}\textsubscript{TH} state. For the high-\textit{V}\textsubscript{TH} state sensing, reduction of the depolarization field is observed in different SiO\textsubscript{2} thicknesses. But due to reduced equivalent oxide thickness (EOT) at a thinner SiO\textsubscript{2}, the required \textit{V}\textsubscript{READ} is also reduced to turn on the channel. Once the channel turns ON, the \textit{E}\textsubscript{FE} saturates at the same value irrespective of the SiO\textsubscript{2} thickness as the \textit{V}\textsubscript{READ} is then screened by the channel. In addition, different non-FE dielectrics are studied, including Al\textsubscript{2}O\textsubscript{3} (supplementary Fig.\ref{fig:sups2}\textbf{a}) and HfO\textsubscript{2} (supplementary Fig.\ref{fig:sups2}\textbf{b}). Again, they confirm that the read disturb-free operation is valid for various non-FE dielectrics, as the feature originates from the device structure.

%% file: bibtex/sections/3_Experimental_verification.tex
\section*{\large Experimental Validation of Read Disturb-Free Operation}

The read disturb-free operation is then validated experimentally on a dual-port FDSOI FeFET. For validation, FeFETs integrated on a 22 nm FDSOI platform are adopted\cite{dunkel2017fefet}. Fig.\ref{fig:fig3}\textbf{a} shows the transmission electron microscopy (TEM) cross-section of the device \cite{dunkel2017fefet}. The device features a 10 nm thick doped HfO\textsubscript{2} as the ferroelectric. By adopting the p-well contact as a dedicated read gate, the FDSOI FeFET intrinsically has dual ports. Detailed process information is described elsewhere\cite{dunkel2017fefet}. First, the device \textit{I}\textsubscript{D}-\textit{V}\textsubscript{G} characteristics for the low-\textit{V}\textsubscript{TH} and high-\textit{V}\textsubscript{TH} states are characterized. Fig.\ref{fig:fig3}\textbf{b} shows the results when sensing the FeFET from either the write gate (WG, i.e., front gate) or the dedicated read gate (RG, i.e., p-well back gate). The \textit{I}\textsubscript{D}-\textit{V}\textsubscript{G} curves sensed on the write gate show a parallel shift in \textit{V}\textsubscript{TH} with respect to different back gate biases for both low-\textit{V}\textsubscript{TH} and high-\textit{V}\textsubscript{TH} states. Therefore, a constant memory window about 1.5 V is observed irrespective of the back gate bias applied. Interestingly, when sensed on the back gate, a constant memory window about 12 V is observed due to the larger equivalent oxide thickness (EOT) of the back gate dielectric\cite{jiang2022asymmetric,mulaosmanovic2021ferroelectric2}. This indicates that through the electrical coupling between the front and back gate, a memory window amplification can be achieved, which opens up the possibility of applying this device for high density storage application, similar to the NAND flash.   


\begin{figurehere}
  \centering
    \includegraphics[width=1\textwidth]{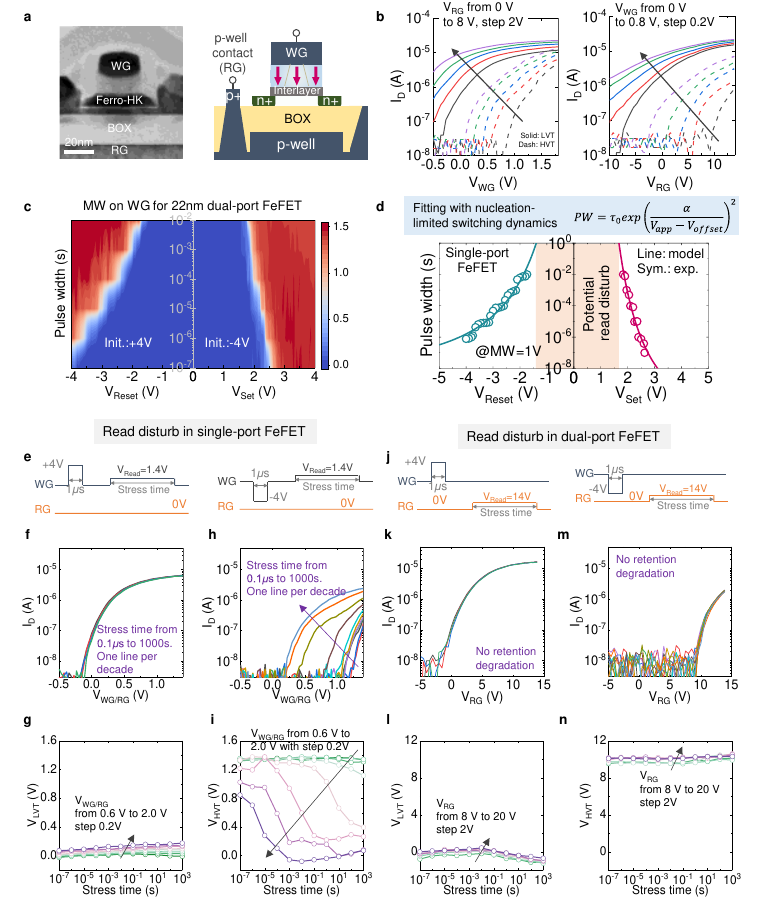}
    \captionsetup{parbox=none} 
    \caption{\textbf{Experimental validation of read disturb-free operation in a dual-port FeFET.} \textbf{a.} TEM \cite{dunkel2017fefet} and schematic cross-sections of a 22 nm high-$\kappa$ metal gate FDSOI FeFET. \textbf{b.} \textit{I}\textsubscript{D}-\textit{V}\textsubscript{G} curves for WG read and RG read measured after $\pm$4 V, 1 $\mu$s write pulses. \textbf{c.} MW map for the FDSOI FeFET using WG for read operation. \textbf{d.} Potential read disturb region is identified in MW contour using the nucleation-limited polarization switching model. \textbf{e.} Waveforms used for read disturb study in the single-port FeFET for both low-\textit{V}\textsubscript{TH} and high-\textit{V}\textsubscript{TH} states. \textbf{f.} No retention degradation is found in the low-\textit{V}\textsubscript{TH} state \textit{I}\textsubscript{D}-\textit{V}\textsubscript{G} curves at \textit{V}\textsubscript{READ}=1.4 V for the stress time up to 1000 s. \textbf{h.} However, the retention for the high-\textit{V}\textsubscript{TH} state is degraded as a function of read voltage and duration. \textbf{g-i.} More read biases are measured to further confirm the read disturb issue. \textbf{j.} Waveforms used for read disturb-free study in the dual-port FeFET. \textbf{k-m.} No retention degradation is observed in both low-\textit{V}\textsubscript{TH} and high-\textit{V}\textsubscript{TH} states. \textbf{l-n.} The retention is robust even for larger \textit{V}\textsubscript{READ} conditions.}
    \label{fig:fig3}
\end{figurehere}

The polarization switching characteristics are also characterized. Fig.\ref{fig:fig3}\textbf{c} shows the memory window (MW) contour map with respect to the write pulse amplitude and pulse width when sensing on the write gate. Two scenarios are considered, one is when the FeFET is initiated to the low-\textit{V}\textsubscript{TH} state with +4 V write pulse and then programmed with a negative reset pulse. The other scenario is opposite where the FeFET is initiated to the high-\textit{V}\textsubscript{TH} state with -4 V write pulse and then programmed with a positive set pulse. It shows a basin around 0 V where no polarization switching is happening. This is desirable as a long retention is required when no voltage is applied. In addition, it also maps out the potential read disturb to the high-\textit{V}\textsubscript{TH} state even when a small positive read pulse is applied. Fig.\ref{fig:fig3}\textbf{d} extracts the required pulse width as a function of pulse amplitude to reach a memory window of 1 V. The dynamics can be well fitted with the nucleation-limited polarization switching model in a thin poly-crystalline film\cite{mulaosmanovic2017switching,deng2020comprehensive}:
\vspace{-2ex}
\begin{equation}
PW=\tau_oexp(\frac{\alpha}{V_{app}-V_{offset}})^2
\end{equation}
where the $\tau_o$, $\alpha$, $V_{app}$, and $V_{offset}$ are the time constant to switch at infinite applied field, a constant representing the switching barrier, applied voltage, and the offset voltage in the film, respectively. It also shows that a small positive read pulse, if continuously applied, can cause disturb to the high-\textit{V}\textsubscript{TH} state.

Next the stability of both low-\textit{V}\textsubscript{TH} and high-\textit{V}\textsubscript{TH} states when stressed with different read voltages are characterized. Fig.\ref{fig:fig3}\textbf{e} shows the waveform applied for single-port operation, where the read gate is grounded all the time and the write pulse and read stress pulse are applied on the write gate. Fig.\ref{fig:fig3}\textbf{f} and \textbf{h} show the \textit{I}\textsubscript{D}-\textit{V}\textsubscript{G} characteristics of the low-\textit{V}\textsubscript{TH} and high-\textit{V}\textsubscript{TH} states, respectively, when a 1.4 V read pulse is stressed from 0.1 $\mu$s to 1000 s. It shows that the low-\textit{V}\textsubscript{TH} state is stable while the high-\textit{V}\textsubscript{TH} state is sensitive to the read stress, confirming the working principles shown in Fig.\ref{fig:Fig2}. In addition, the extracted \textit{V}\textsubscript{TH} for the low-\textit{V}\textsubscript{TH} and high-\textit{V}\textsubscript{TH} states are shown in Fig.\ref{fig:fig3}\textbf{g} and \textbf{i}, respectively. Different read voltages ranging from 0.6 V to 2.0 V are applied. It clearly shows that low-\textit{V}\textsubscript{TH} state is stable while the high-\textit{V}\textsubscript{TH} state can be destroyed if a long enough read bias is applied, even with a small read voltage. This disturb is avoided in dual-port FeFET. Fig.\ref{fig:fig3}\textbf{j} shows the waveform applied for verification. The write pulse is applied on the write gate while read voltage ranging from 8 V to 20 V is applied on the read gate to stress the device. The \textit{I}\textsubscript{D}-\textit{V}\textsubscript{G} characteristics of the low-\textit{V}\textsubscript{TH} and high-\textit{V}\textsubscript{TH} states are shown in Fig.\ref{fig:fig3}\textbf{k} and \textbf{m}, respectively, when a 14 V read pulse is stressed from 0.1 $\mu$s to 1000 s. No degradation is observed. The extracted \textit{V}\textsubscript{TH} for the low-\textit{V}\textsubscript{TH} and high-\textit{V}\textsubscript{TH} states are shown in Fig.\ref{fig:fig3}\textbf{l} and \textbf{n}, respectively. Even with the read voltage of 20 V, no degradation in \textit{V}\textsubscript{TH} is observed. These results therefore confirm that the read operation incurs no disturb to the polarization in a dual-port FeFET.

%% file: bibtex/sections/4_Applications.tex
\section*{\large Applications of Dual-Port FeFETs}

With the read disturb-free feature and the separated write and read paths, many interesting applications are enabled with dual-port FeFET, as discussed in the introduction. In this section, two applications are experimentally demonstrated, including FPGA leveraging the read disturb-free feature and tunable analog electronics exploiting separated write and read paths taking frequency tunable ring oscillator as an example. In a conventional FPGA, static random access memory (SRAM) is typically used to construct look up table (LUT) in the configurable logic block (CLB) and routing switches in the connection box (CB) and switch box (SB), which has a low density and high leakage. Using FeFET for LUT and routing switches, compact FPGA can be achieved\cite{yu2022hardware,chen2018power,xu2022ferroelectric}, as depicted in Fig.\ref{fig:Fig4}\textbf{a}. Being nonvolatile and having a transistor structure, FeFET makes a compact nonvolatile switch, outperforming many solutions combining memory and transistor together\cite{yu2022hardware}. However, as a routing switch, the configuration is stored in the FeFET and a constant \textit{V}\textsubscript{READ} between the low-\textit{V}\textsubscript{TH} and high-\textit{V}\textsubscript{TH} needs to be applied to activate the configuration. In this case, there is a concern over state stability in a single-port FeFET while the dual-port FeFET could ensure continual operation without disturb to polarization. 

\begin{figurehere}
  \centering
    \includegraphics[width=1\textwidth]{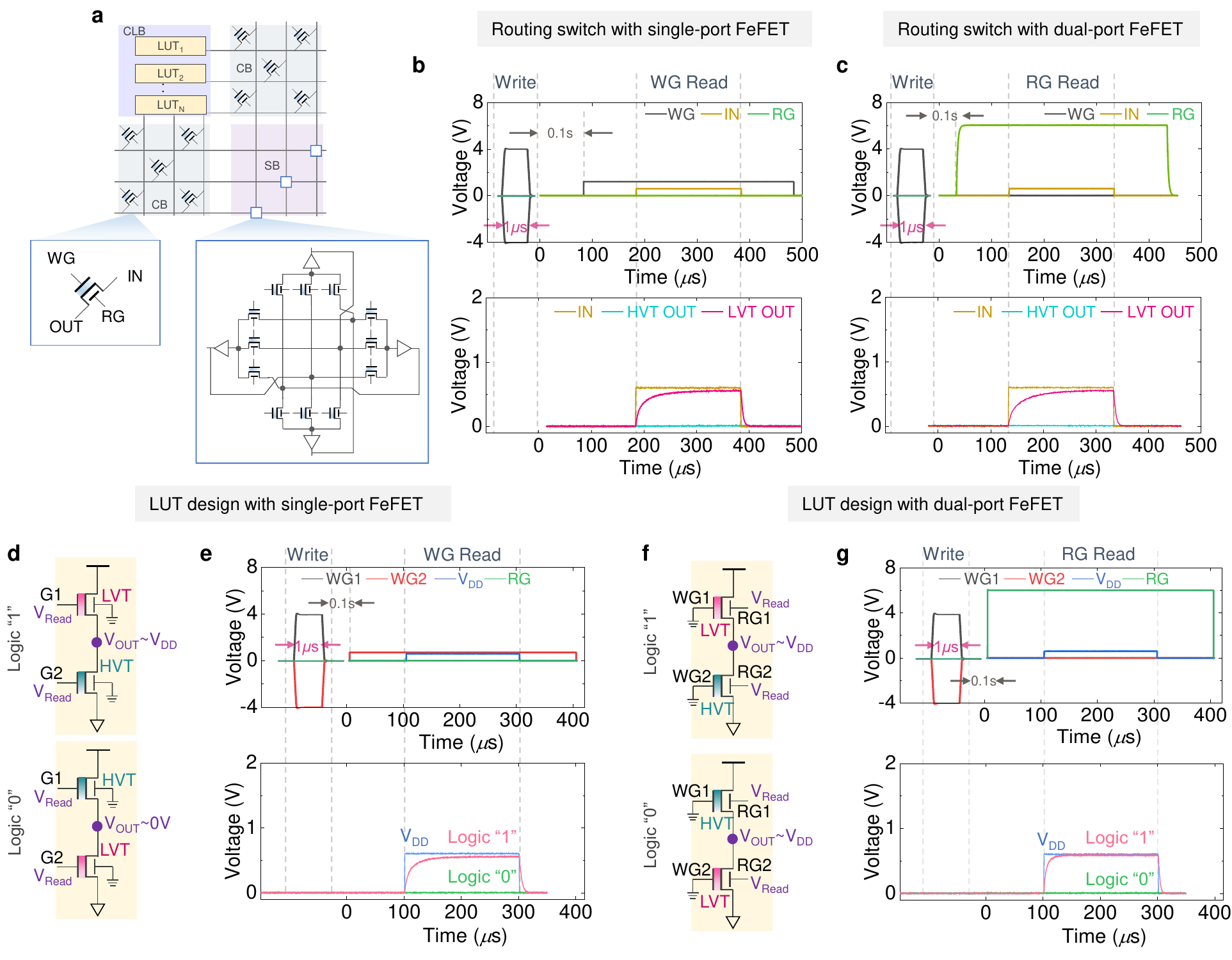}
    \captionsetup{parbox=none} 
    \caption{ \textbf{Exploiting the read disturb-free feature of dual-port FeFET for FPGA design.} \textbf{a.} Compact FPGA using FeFET for LUT (CLB) and routing switches (CB, SB). \textbf{b.} The transient waveforms of the single-port FeFET based routing switch for the experimental verification of the block/pass configurations and activation. \textbf{c.} The similar waveforms for the dual-port FeFET based routing switch. Activation is done by the RG. \textbf{d.} A LUT cell consisting of two FeFETs with complementary \textit{V}\textsubscript{TH} states. \textbf{e.} The experimental transient waveforms of the single-port FeFET based LUT for functionality verification. \textbf{f.} A dual-port FeFET based LUT cell. \textbf{g.} The waveforms of the dual-port FeFET based LUT. Logic “1” and “0” can be obtained for both single- and dual-port configurations, confirming the feasibility of replacing the single-port FeFET with the dual-port FeFET for read disturb-free operations.}
    \label{fig:Fig4}
\end{figurehere}

In the following, the LUT and routing switch functionalities using both discrete single-port and dual-port FeFETs are verified. A single FeFET cell can function as a routing switch, as shown in Fig.\ref{fig:Fig4}\textbf{a}. When configuring a FeFET into low-\textit{V}\textsubscript{TH}/high-\textit{V}\textsubscript{TH} state, the switch will be passing/blocking the incoming signal, respectively. The feasibility of passing or blocking input for a single-/dual-port FeFET is verified with programming on the WG and activating the configuration on WG/RG, respectively. Fig.\ref{fig:Fig4}\textbf{b} shows when the FeFET is programmed to the high-\textit{V}\textsubscript{TH}/low-\textit{V}\textsubscript{TH} state, activating the configuration from WG in a single-port FeFET could block/pass the input signal, respectively. This is because the channel is OFF for high-\textit{V}\textsubscript{TH} state while turns ON for low-\textit{V}\textsubscript{TH} state. For a dual-port FeFET, configuration programming is conducted on the WG while activation is performed on the RG. Fig.\ref{fig:Fig4}\textbf{c} shows the routing switch function in a dual-port FeFET. It shows a similar function as that in a single-port FeFET, where the input signal is blocked/passed to the output when the FeFET is in the high-\textit{V}\textsubscript{TH}/low-\textit{V}\textsubscript{TH} state, respectively. Therefore, these results confirm successful routing function in a dual-port FeFET. Along with its read disturb-free feature, dual-port FeFETs can realize a robust routing switch.

A LUT cell consists of two FeFETs with complementary \textit{V}\textsubscript{TH} states, as shown in Fig.\ref{fig:Fig4}\textbf{d}. The logic information can be encoded as the stored complementary \textit{V}\textsubscript{TH} states and encoded information can be easily read out as the internal node voltage. For example, to store logic “1”, the upper/lower FeFET is programmed to the low-\textit{V}\textsubscript{TH}/high-\textit{V}\textsubscript{TH} state, respectively. To read the stored information, a read signal is applied to WG in a single-port FeFET as shown in Fig.\ref{fig:Fig4}\textbf{d}, and RG in a dual-port FeFET as shown in Fig.\ref{fig:Fig4}\textbf{f}, respectively. The waveforms and measured results for single-port and dual-port FeFET based LUT cell are shown in Fig.\ref{fig:Fig4}\textbf{e} and \textbf{g}, respectively. It confirms that for both single-port and dual-port FeFET, the output is high for logic '1' and low for logic '0', thus confirming the LUT cell operation. Combining with the routing switch to implement a CB and SB, a complete solution based on dual-port FeFET can enable highly efficient and robust FPGA design.

As an example of tunable analog electronics that can benefit from separated write and read paths in a dual-port FeFET, a nonvolatile ring oscillator is demonstrated. Due to setup limitation, a 3-stage ring oscillator is tested where the first stage consists of a dual-port FeFET and a resistor while CMOS inverters are used for the second and third stages, as shown in Fig.\ref{fig:Fig5}\textbf{a}. The WG is used to program the FeFET \textit{V}\textsubscript{TH} to adjust the oscillation frequency. The RG then can participate in signal transmission without the need of analog multiplexer. Oscillation waveforms are shown for FeFET written with 4 V and 2.5 V in Fig.\ref{fig:Fig5}\textbf{b}. The oscillation frequency reduces from 121 kHz to 111 kHz as the \textit{V}\textsubscript{TH} increases due to a larger discharge delay in the first stage. Furthermore, a 5-stage dual-port FeFET based ring oscillator is simulated, as shown in Fig.\ref{fig:Fig5}\textbf{c}. The oscillation frequency presented in Fig.\ref{fig:Fig5}\textbf{d} decreases as the \textit{V}\textsubscript{TH} increases, consistent with the experiment. Moreover, thanks to more FeFETs used, a wide frequency tuning range is achieved. It is also expected that replacing the resistor with PMOS can retain the tuning capability while also eliminating the DC power consumption, making the dual-port FeFET ring oscillator a promising approach for frequency adjustment.

\begin{figurehere}
  \centering
    \includegraphics[width=0.9\textwidth]{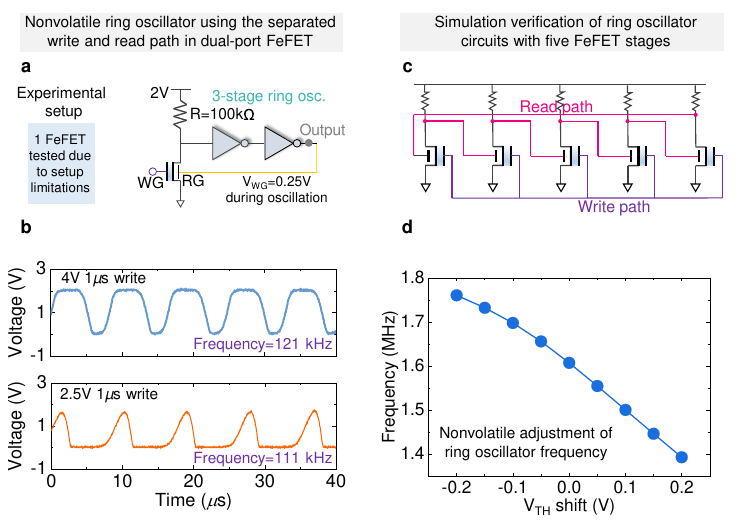}
    \captionsetup{parbox=none} 
    \caption{ \textbf{Leveraging the separated write and read paths in dual-port FeFETs for frequency tunable oscillator.} \textbf{a.} Experimental setup of 3-stage ring oscillator. \textbf{b.} Measured oscillation waveform show tunability of frequency with FeFET \textit{V}\textsubscript{TH} programming. The lower the \textit{V}\textsubscript{TH}, the higher the oscillation frequency. \textbf{c.} The schematic for the 5-stage ring oscillator. \textbf{d.} Simulated results show the ability of a wider tuning range with more dual-port FeFETs in the oscillator.}
    \label{fig:Fig5}
\end{figurehere}

%% file: bibtex/sections/5_Conclusion.tex
\vspace{-4ex}
\section*{\large Conclusion}
We have performed a comprehensive evaluation of the read disturb-free operation in dual-port FeFETs through combined experimental characterization and theoretical analysis. We have shown that due to its unique structure, the read disturb-free feature is intrinsic to the dual-port FeFETs and can be mainatained with the non-ferroelectric dielectric scaling for read voltage reduction. This design addresses a key challenge in the single-port FeFETs in applications that need a long read stress. We have also explored the applications, including FPGA and tunable analog electronics (e.g., ring oscillator) that can benefit from the read disturb-free feature and separated write and read paths. Through this work, it is clear that another useful device can be added into the ferroelectric device families and enrich their versatility, thereby empowering applications calling for diverse functionalities.   

\vspace{-4ex}

%% file: bibtex/sections/Supplementary.tex
\renewcommand{\thefigure}{S\arabic{figure}}
\renewcommand{\thetable}{S\arabic{table}}
\setcounter{figure}{0}
\setcounter{table}{0}

\newpage
\begin{center}
\title{\textbf{\Large Supplementary Materials}}
\end{center}

\vspace{-4ex}
\section*{\large Device Fabrication}

In this paper, the fabricated fully depleted silicon-on-insulator (FDSOI) ferroelectric field effect transistor (FeFET) features a poly-crystalline Si/TiN/doped HfO\textsubscript{2}/SiO\textsubscript{2}/p-Si gate stack. The devices were fabricated using a 22 nm node gate-first high-$\kappa$ metal gate CMOS process on 300 mm silicon wafers. The buried oxide is 20nm SiO\textsubscript{2}. Detailed information can be found in \cite{dunkel2017fefet, mulaosmanovic2021ferroelectric}. The ferroelectric gate stack process module starts with growth of a thin SiO\textsubscript{2} based interfacial layer, followed by the deposition of the doped HfO\textsubscript{2} film. 
A TiN metal gate electrode was deposited using physical vapor deposition (PVD), on top of which the poly-Si gate electrode is deposited. The source and drain n+ regions were obtained by phosphorous ion implantation, which were then activated by a rapid thermal annealing (RTA) at approximately 1000 $^\circ$C. This step also results in the formation of the ferroelectric orthorhombic phase within the doped HfO\textsubscript{2}.

\newpage
\section*{\large Electrical Characterization}

The experimental verification was performed with a Keithley 4200-SCS Semiconductor Characterization System (Keithley system), a Tektronix TDS 2012B Two Channel Digital Storage Oscilloscope (oscilloscope), and a Keysight 81150A Pulse Function Arbitrary Generator (waveform generator). Two 4225-PMUs (pulse measurement units) were utilized to generate proper waveforms. In the experimental characteristics of the 22 nm FDSOI FeFET, all signals were generated by the Keithley system. The drain currents were also capaturd by the Keithley system. The \textit{V}\textsubscript{TH} was extracted with a constant current of $I_D=10^{-7} W/L $ \textit{A}. All signals in FPGA experimental verification were generated by Keithley system. The output pulses were captured by the oscilloscope. In the experimental verification of the ring oscillator, the FeFET, two inverters, and the resistor were connected externally on a breadboard. The write operations on the FeFET were provided by the Keithley system. The \textit{V}\textsubscript{DD} for the inverters was given by the waveform generator. The output pulses were captured by the oscilloscope. 

\newpage
\section*{\large Band Diagrams to Show Read Disturb-Free Operation for FeFET Written to Low-\textit{V}\textsubscript{TH} State}

\begin{figurehere}
  \centering
    \includegraphics[width=0.75\textwidth]    {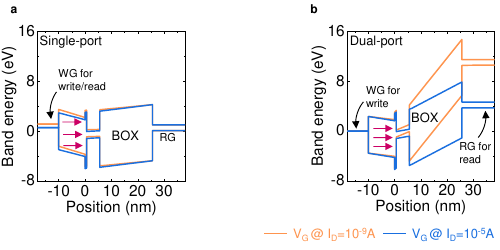}
    \captionsetup{parbox=none} 
    \caption{ \textbf{Low-\textit{V}\textsubscript{TH} state band diagrams during read operation for the single- and dual-port FeFET.} \textbf{a.} For the low-\textit{V}\textsubscript{TH} state in the single-port FeFET, \textit{E}\textsubscript{Dep} decreases when \textit{V}\textsubscript{READ} increases. \textbf{b.} For the same state in the dual-port FeFET, there is no change in \textit{E}\textsubscript{Dep} when \textit{V}\textsubscript{READ} increases.}
    \vspace{-4ex}
    \label{fig:sups1}
\end{figurehere}
In contrast to the read operation of the high-\textit{V}\textsubscript{TH} state metioned before, during the read operation for the low-\textit{V}\textsubscript{TH} state in Fig.\ref{fig:sups1}\textbf{a}, the applied electric field, \textit{E}\textsubscript{App}, caused by the read voltage has the same direction as the polarization field of the ferroelectric layer. This enhances the ferroelectric polarization and reduces the de-polarization field, \textit{E}\textsubscript{Dep}. Therefore, no retention degradation occurs in the experimental verification for the low-\textit{V}\textsubscript{TH} state FeFET. For the dual-port FeFET in low-\textit{V}\textsubscript{TH} state in Fig.\ref{fig:sups1}\textbf{b}, the applied read voltage drops on the non-FE layer. Although the \textit{E}\textsubscript{App} has the same direction as the \textit{E}\textsubscript{Dep}, the formed channel screens the \textit{E}\textsubscript{App}, providing a robust retention.

\newpage
\section*{\large Impact of Non-Ferroelectric Dielectric Materials on Read Disturb-Free Feature}
\begin{figurehere}
  \centering
    \includegraphics[width=1\textwidth]{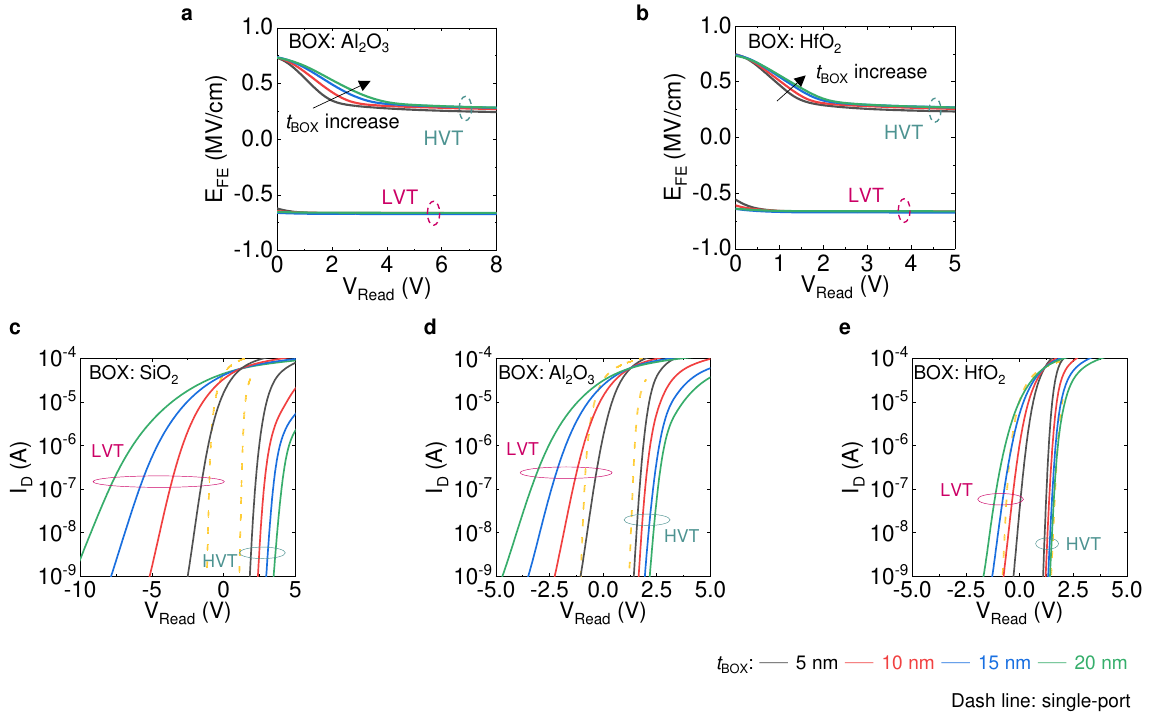}
    \captionsetup{parbox=none} 
    \caption{ \textbf{Impact of different types and thicknesses of non-ferroelectric materials} Ferroelectric field with respect to different read biases and BOX thicknesses for \textbf{a.} Al\textsubscript{2}O\textsubscript{3} and \textbf{b.} HfO\textsubscript{2} BOX. HfO\textsubscript{2} BOX exhibts smaller V\textsubscript{READ} for the FeFET to be inverted and form the channel layer. \textbf{c-e.} show the \textit{I}\textsubscript{D}-\textit{V}\textsubscript{G} curves for different BOX materials.}
    \label{fig:sups2}
\end{figurehere}
More TCAD simulations on different BOX materials and thicknesses are performed. Fig.\ref{fig:sups2}\textbf{a} and \textbf{b} show the \textit{E}\textsubscript{FE} for Al\textsubscript{2}O\textsubscript{3} and HfO\textsubscript{2}, respectively. Due to the high dielectric constant of HfO\textsubscript{2}, the equivalent oxide thickness (EOT) is reduced. Therefore, the required \textit{V}\textsubscript{Read} to turn on the channel is also reduced (Fig.\ref{fig:sups2}\textbf{b}). Similar to the case of SiO\textsubscript{2} BOX mentioned before, the \textit{E}\textsubscript{FE} saturates at the same value irrespective of the thickness after the channel turns on as the \textit{V}\textsubscript{Read} is then screened by the channel.
Fig.\ref{fig:sups2}\textbf{c-e} show the \textit{I}\textsubscript{D}-\textit{V}\textsubscript{G} curves for SiO\textsubscript{2}, Al\textsubscript{2}O\textsubscript{3} , and Hf\textsubscript{2} BOX. For the same thickness, the  HfO\textsubscript{2} BOX exhibits smaller \textit{V}\textsubscript{READ} to achieve similar drain current.